Title: Nonradiative DKR processes: revisiting the theory. III. Multimode approaches
Authors: Mladen Georgiev (1) and Fabio DeMatteis (2) (Institute of Solid State
   Physics Bulgarian Academy of Sciences, 1784 Sofia Bulgaria (1) and Dipartimento
   di Fisica, II Universita degli Studi di Roma "Tor Vergata", 00133 Roma Italia (2))
Comments: 16 pages including 3(5) tables and 4(5) figures, all pdf format
Subj-class: cond-mat

We outline a method for dealing with the relaxation of optically excited NaI F centers in terms of a single effective frequency along a multimode coordinate. The 2s-2p mixing through coupling to a $T_{1u}$ vibrational mode is also discussed with optimistic conclusions. Equilibrium nonradiative deexcitation rates are estimated and compared with intervibrational relaxations so as to assess the efficiency of the nonequilibrium DKR deexcitation. The estimates show the former not to be competitive, due to the large number of phonons exchanged during the relaxation.

1. Introduction

The single-frequency uni-mode oscillator model of Part I does not match the experimental situation quite well. As a matter of fact, observed vibrational frequencies of what is believed to be the breathing mode vibrations centered at an anion site are consent to decreasing in the order $\omega_{LO} > \omega_\alpha > \omega_{F*} > \omega_F$. First, we see that the mode frequency coupled to an F center in the electronic excited state is superior to the one coupled to its ground state. Second, the vibration around an empty vacancy is of higher frequency than the one around an F center which suggests renormalization by the electron-mode interaction in that the spring constant is the lower the more compact the electron cloud seated in and around the cavity. Third, although the vibrational frequencies coupled to the electronic states of the F center in NaI are not precisely known, there are some indirect estimates of the related spring constants [1] and frequencies [2] coupled to the electronic transitions. For this variety of reasons extending our arguments to a bi-frequency model seems appropriate.

As shown by Christov [3], one can still use single-frequency mathematics by introducing an effective frequency when more than one vibrational mode is known to couple to an electronic state,

$$\omega_{eff}^{-1} = \sum_i \omega_i^{-1} E_{Ri} / \sum_i E_{Ri} = \sum_i \omega_i^{-1} E_{Ri} / E_R, \qquad (1)$$

where the subscripts to $\omega_i$ refer to the different components and $E_{Ri}$ are their respective reorganization energies, $E_R = \sum_i E_{Ri}$ is the total reorganization energy. This theory is not directly applicable to our case where two electronic states are involved, each coupled to a different mode rather than one electronic state coupled to two modes. However we can use its basis to formulate a working effective frequency approach to the real problem. The difficulty is in matching the vibronic energies in ground and excited electronic F states for deriving a transition rate.

Spectroscopic studies have portrayed the structure of the first excited F center state as a vibronic mixture of 2s- and 2p- like components [4]. Mixing occurs along the relaxation path somewhat away from the initial configuration at which absorption has taken place. As a result, the *2s*-like state lying initially above the *2p*-like state, it plunges below it after mixing. The equilibriums of the mixed components are attained at nearly the same configuration with a mostly *2s*-like character lying lower than a mostly *2p*-like character. Spectroscopists deduce a longer radiative lifetime following one-photon absorption and a shorter radiative lifetime following a two-photon absorption. Inasmuch as $|2s\rangle$ and $|2p\rangle$ are of the opposite parities, they should mix by coupling to an odd-parity vibration. The problem has been studied earlier by Ham et al. [5] using Pseudo-Jahn-Teller arguments and later by Martinelli et al. [6] by means of recursion methods. Yet, it is tempting to explore the coupling to the odd $T_{1u}$ mode of the six cations nearest-neighboring the anion vacancy.

We consider the situation in which the predominantly 2s-like mixed character (2s') couples to two different vibrational modes. In so far as the 2s-like component is compact as is the 1s-like state, we presume that both s-states will couple to the same $A_{1g}$ symmetry breathing-mode vibration of the six nn cations. Consequently, the foregoing single-frequency analysis will be applicable to the 1s-2s interrelationship at $\omega = \omega_{1s} = \omega_{2s}$, while the frequency $\omega_{2p}$ coupled to the 2p-like state may be higher. Its mode symmetry will be subject to further considerations. The effective frequency coupled to 2s' is given by equation (1). It yields $\omega_{eff} \sim \omega_{1s}$ if $E_{R2s} \gg E_{R2p}$. Under these conditions the formal analysis of Part I will apply to the system, composed of the 1s-like state and the 2s' mixed character in which no change in the effective vibrational frequency occurs during the nonradiative deexcitation transition.

In what follows, we shall pursue these premises subsequently. Previous configurational coordinate calculations using the semi-continuum electronic potential show that breathing mode coupling alone grossly underestimates the electron-mode coupling strength in 2p-state of the F center [1]. A reasonable fit to experimental optical energies has only been achieved by largely increasing the coupling constant and reversing its sign. Both changes are taken to signify a parallel coupling to the $A_{1g}$ lattice mode, due to a partial delocalization of the 2p-like state. This delocalization results from the mixing with a 2p-like bound-polaron state, the bound polaron connection being suggested earlier [7]. Now if both the local breathing-mode (LBM) vibration and the $A_{1g}$ phonon mode (PM) couple to the 2p-like character, the effective vibrational frequency (~0.012eV) resolves in two respective components as $\omega_{eff} = 1.5 \times \omega_{LBM}\omega_{PM} / (\omega_{LBM}+\omega_{PM})$ with $\omega_{LBM}=0.0129$eV and $\omega_{PM}=0.0228$eV. The LO-coupling constant of the delocaized 2p-state, otherwise obtained by semi-continuum potential calculations, should be magnified by a 'polaron factor' $g_p = -\omega_{LO} \times \sqrt{(M_{LO}v_a/4\pi\varepsilon_p e^2)}$, $v_a$ is the unit-cell volume and $\varepsilon_p = 1/(1/\varepsilon_0+1/\varepsilon_\infty)$, to reflect the fact that the coupling is to the whole lattice rather than to just a local mode [7]. From the local value $b_{2p2p} = 2.69$ meV/Å we obtain the Table I data (-1.071eV/Å at $g_p = -399$).

2. Multimode coupling

We consider the coupling to a manifold of configurational coordinates $q_i$ of two diabatic potentials, e.g.

$$V_1(q) = \sum_i \tfrac{1}{2} K_i q_i^2 \equiv \tfrac{1}{2} K q_r^2$$

$$V_2(q) = \sum_i \tfrac{1}{2} K_i (q_i - q_{i0})^2 + Q \equiv \tfrac{1}{2} K (q_r - q_{r0})^2 + Q$$

where $q_{i0}$ are the respective equilibrium positions. We postulated a diagonal reaction coordinate $q_r$ by way of $q_r^2 = \sum_i q_i^2$. Defining the partial reorganization energies

$$E_{Ri} = \tfrac{1}{2} K_i q_{i0}^2$$

we sum up to get

$$E_R = \sum_i E_{Ri} = \sum_i \tfrac{1}{2} K_i q_{i0}^2 = \tfrac{1}{2} K q_{r0}^2.$$

In the effective-mode picture, $E_R$ appears as total reorganization energy along $q_r$. Incorporating partial and total reorganization energies, we arrive at

$$1/K = \sum_i (1/K_i)(E_{Ri}/E_R)$$

which defines the effective force constants along $q_i$. Once K has been found, we set $q_{i0} = -G_i/K_i$, $q_{r0} = -G/K$ in $q_{r0}^2 = \sum_i q_{i0}^2$ to get

$$(G/K)^2 = \sum_i (G_i/K_i)^2$$

which defines the electron-mode coupling constant along $q_i$.

Introducing dimensionless mode coordinates $\xi_i = \sqrt{(K_i / \hbar\omega_i)}\, q_i$, the diabatic potentials transform to

$$V_1(\xi) = \sum_i \tfrac{1}{2} \hbar\omega_i \xi_i^2 \equiv \tfrac{1}{2} \hbar\omega \xi_r^2$$

$$V_2(\xi) = \sum_i \tfrac{1}{2} \hbar\omega_i (\xi_i - \xi_{i0})^2 + Q \equiv \tfrac{1}{2} \hbar\omega (\xi_r - \xi_{r0})^2 + Q$$

where $\hbar\omega$ and $\hbar\omega_i$ substitute for K and $K_i$, respectively. Now, the foregoing analysis yields the effective vibrational frequency along $\xi_r$:

$$1/\omega = \sum_i (1/\omega_i)(E_{Ri}/E_R)$$

We see that introducing the reaction coordinate is fruitful in cases where it reduces a multi-mode problem to a single-mode one.

### 3. 2s-2p phonon-mediated mixing

In Part I, we calculated diabatic potentials for the F center states in NaI following the method outlined therein. The electron-mode coupling operator was taken in a local breathing-mode form. The calculated parameters of configurational-coordinate (CC) diagrams were listed in Table I, while resulting F center CC diagrams were depicted in Figure 1 *loc.cit*.

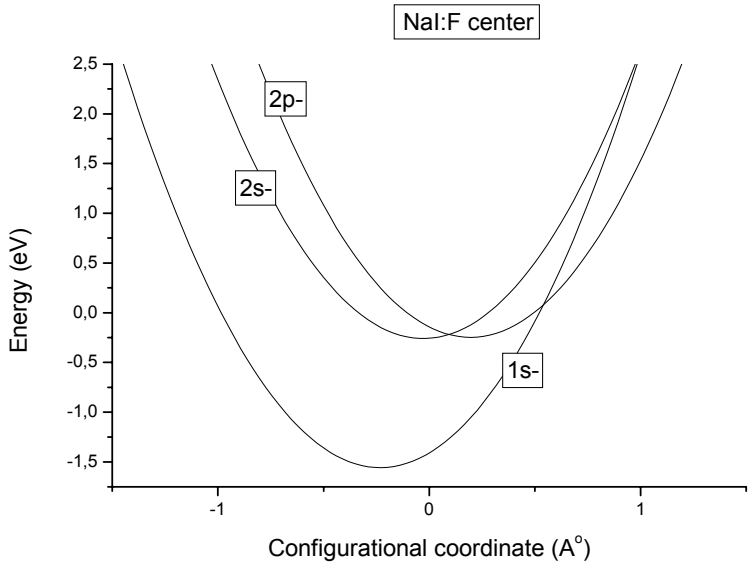

Figure 1:
Diabatic potentials of the NaI F center as generated by the 1s-, 2s-, and 2p- like states using Table I data. The 2p- like state, now partially delocalized and coupled to the $A_{1g}$ lattice mode, has its parabola shifted to the right along the breathing-mode coordinate ($A_{1g}$-symmetry too), as compared to Figure 1 of Part I.

We now extend the scheme whereby the adiabatic splitting of the vibronic potentials is regarded as mixing effected by the coupling of the 2s- and 2p- like excited F center states in NaI to one another mediated by a vibrational mode. As suggested in Section 1, the coupled vibration is a bimode mixture expected to occur at the vibrational mode frequency. In accordance, a single mode coordinate q is assumed.

The mixing is of two 'diabatic parabolae', correspondingly,

$V_{2s}(q) = \frac{1}{2}Kq^2 + b_{2s2s}q + E_{2s}$

$V_{2p}(q) = \frac{1}{2}Kq^2 + b_{2p2p}q + E_{2p}$ (2)

which avoid crossing at $q = q_c$ by virtue of their coupling energy $2V_{2s2p}$. There are two interrelated ways of effecting a 2s-2p coupling: (i) through the mediation of a vibrational mode

$V_{2s2p} = <2s|b(r)|2p>$,

where b(r) is the electron-mode coupling operator, or (ii) by tunneling.

By mixing equations (2) we arrive at

$E_{\pm}(q) = \frac{1}{2}\{V_{2p}(q)+V_{2s}(q) \pm \sqrt{[(V_{2p}(q)-V_{2s}(q))^2 + 4|V_{2s2p}|^2]}\}$

$$= \tfrac{1}{2}\{2Kq^2+E_{2p}+E_{2s} \pm \sqrt{[((b_{2p2p}-b_{2s2s})q+E_{2p}-E_{2s})^2 + 4|V_{2s2p}|^2]}\}$$

$$\equiv Kq^2 \pm \tfrac{1}{2}\sqrt{[(2Gq+D)^2 + 4|V_{2s2p}|^2]} + \tfrac{1}{2}(E_{2p}+E_{2s}) \tag{3}$$

setting $2G = -(b_{2p2p}-b_{2s2s})$, $D = -(E_{2p}-E_{2s})$. From Table I we have $G = 0.623$ eV/Å, $D = -0.114$ eV. The strong-mixing criterion being $E_R > Q$, we get from the definitions in Part I: $b_{2p2p} \times (-2G/K) > D$ which is met by the Table I data.

Now, because of the even parity of the breathing mode ($A_{1g}$), the matrix element $V_{2s2p}$ will be vanishing for that mode, as shown in Table I. As a result, the adiabatic energies equation (3) will decompose back into the original diabatic energies equations (2). We stress that the mixing-mediating vibration should have at least one signifying odd-parity component in order to give rise to adiabatic energies of the equation (3) type.

For a better correspondence with the experimental emission data, we next assume partial coupling to the $A_{1g}$ lattice mode for the 2p- like state. This magnifies the absolute coupling coefficient by the polaron factor $g_p$, as explained above, and results in shifting the 2p-like parabola to the other side of the origin, as in Figure 1 herein.

### 4. 2s-2p vibronic mixing

The hypothesis of a vibronic mixing of the lowest-lying excited F center states has been put forward and explored by Ham [5] with a less than affirmative result as regards the $T_{1u}$ coupling. We now revive the debate in view of the subsequent developments [6]. In particular, we refer to the so-called Asymmetric Pseudo-Jahn-Teller Effect (APJTE) [8,9], as described by the Hamiltonian

$$H = \tfrac{1}{2}(\sum_i \mathbf{P}_i^2/M_i + \sum_i K_i q_i^2)$$

$$+ \tfrac{1}{2}E_{sp}(\sum_i |p_i\rangle\langle p_i| - |s\rangle\langle s|) + \sum_i (G_i q_i + D_i)(|p_i\rangle\langle s| + |s\rangle\langle p_i|) \tag{4}$$

where $q_i$ ($i = x,y,z$) are the three $T_{1u}$ coordinates at the cubic $O_h$ symmetry F center site, $E_{sp}$ is the *2p-2s* energy gap at the non-mixing configuration $q_i = -D_i/G_i$. $\mathbf{P}_i$ ($\mathbf{P}^2 = \sum_i \mathbf{P}_i^2$), $M_i$, and $K_i$ are the mode component momenta, masses, and spring constants, respectively, $G_i$ are the mixing constants and $D_i$ are assymetry parameters. In view of the isotropic space along the coordinate axes, we can set $K_i = K$, $M_i = M$, $G_i = G$, $D_i = D$. The assymetry $D$ is a static mixing term which adds to the mode-mixing energy $Gq$. The traditional symmetric Pseudo-Jahn-Teller Effect (PJTE) Hamiltonian obtains from (8) at $D = 0$.

Diagonalizing (1) in the $\{|2s\rangle, |2p_i\rangle\}$ basis at $\mathbf{P} = \mathbf{0}$ we obtain the following adiabatic energy eigenvalues

$$E_\pm(q) = \tfrac{1}{2}\{\sum_i K_i q_i^2 \pm [\sum_i (2G_i q_i + D_i)^2 + E_{sp}^2]^{1/2}\}$$

$$= \tfrac{1}{2}\{Kq^2 \pm [\sum_i (2Gq_i + D)^2 + E_{sp}^2]^{1/2}\} \tag{5}$$

with $q^2 = \Sigma_i q_i^2$. We stress the similarity of equations (5) to the foregoing equations (3). It can be seen that strongly coupled PJTE potentials ultimately transform into weakly coupled ones as the assymetry energy D is increased. In particular, a moderate D-value is characteristic of the calculated 2s' and 2p' F center adiabatic energies in NaI, as implied by Figure 2. We conclude that vibronic 2s-2p mixing combined with a static component is likely to account for the character of the excited F center state in NaI and possibly other alkali halides.

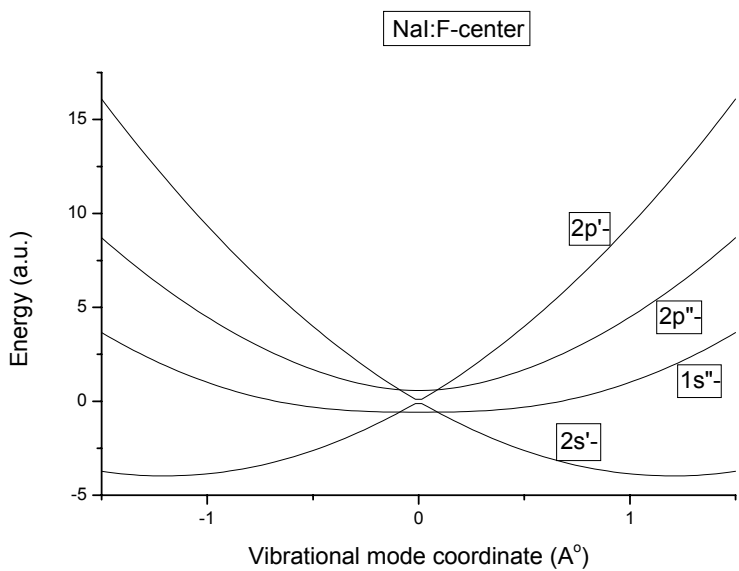

Figure 2:

Adiabatic potentials 2s'- & 1s"- (lower branches) and 2p'- & 2p"- (upper branches) for mixing the 2s- & 2p- and 1s & 2p- like states, respectively, of the NaI F center by $T_{1u}$-mode coupling (symmetric pseudo-Jahn-Teller effect at D=0). These are calculated using Table III data relative to the unmixing configuration and the energy reference midway between the lower and upper branches. They show the 1s-2p mixing to be weak and the 2s-2p mixing to be strong.

The respective experimental optical energies and calculated adiabatic energy parameters of the F center states in NaI are all listed in Table II. The crucial point for the PJTE hypothesis is the reliability of the obtained static 2s-2p mixing energy. From (5) the 2p-2s energy gap at q = 0 is $(D^2 + E_{sp}^2)^{1/2}$; we interpret D as the $\langle 2p|H_{AD}|2s\rangle$ coupling energy where $H_{AD}$ is the adiabatic Hamiltonian obtained from (4) at **P** = **0**. We also note that coupling of the two level system to an external field **F** will add up extra energy of the form -**p**·**F** to D where **p** is the mixing dipole **p** = $\langle 2p|e\mathbf{r}|2s\rangle$. An example of this type is the Stark effect splitting of excited-state 2p-levels in an external electric field.

The mixed eigenstates corresponding to the adiabatic energy eigenvalues (5) are of importance too. They read [10]:

$$\Phi_+(r;q) \equiv |2p'\rangle = 2^{-\frac{1}{2}}\{[\cos(\phi(q)/2)-\sin(\phi(q)/2)]|2s\rangle + [\cos(\phi(q)/2)+\sin(\phi(q)/2)]|2p\rangle\} \quad (6)$$

$$\Phi_-(r;q) \equiv |2s'\rangle = 2^{-\frac{1}{2}}\{[\cos(\phi(q)/2)+\sin(\phi(q)/2)]|2s\rangle - [\cos(\phi(q)/2)-\sin(\phi(q)/2)]|2p\rangle\}$$

where

$$\phi(q) = \tan^{-1}(E_{sp}/\sqrt{\sum_i(2G_iq_i+D_i)^2}) \quad (7)$$

and $|p\rangle$ standing for the linear combination $|p\rangle = \sum_i \alpha_i |p_i\rangle$ with $\alpha_i = q_i/q$, the components of a unit vector along the displacement $q = (q_x, q_y, q_z)$ [5]. Clearly, $\phi(q) = \frac{1}{2}\pi$ for $q_i = -D_i/2G_i$ which gives $|2p'\rangle = |2p\rangle$ and $|2s'\rangle = |2s\rangle$. Note that $q_i = -D_i/2G_i$ is the high symmetry (cubic) configuration at $D_i = 0$. We also see that away from the latter configuration at large $q_i$: $|2p'\rangle = 2^{-\frac{1}{2}}(|2s\rangle + |2p\rangle)$ and $|2s'\rangle = 2^{-\frac{1}{2}}(|2s\rangle - |2p\rangle)$, the symmetric and anti-symmetric linear combinations of the basis functions.

Following Ham [5], we next introduce

$$\lambda(q) = [\sin(\phi(q)/2) - \cos(\phi(q)/2)] / [\sin(\phi(q)/2) + \cos(\phi(q)/2)], \quad (8)$$

the admixed fraction of $|2p\rangle$ state in $|2s'\rangle$ and $|2s\rangle$ state in $|2p'\rangle$, to get

$$|2p'\rangle = (1+\lambda^2)^{-\frac{1}{2}}\{|2p\rangle - \lambda|2s\rangle\}$$

$$|2s'\rangle = (1+\lambda^2)^{-\frac{1}{2}}\{|2s\rangle + \lambda|2p\rangle\} \quad (9)$$

Using these, the radiative lifetimes from $|2p'\rangle$ and $|2s'\rangle$ are:

$$t_p \sim \langle 1s|er|2p'\rangle^{-2} = (1+\lambda^2)\langle 1s|er|2p\rangle^{-2}$$

$$t_s \sim \langle 1s|er|2s'\rangle^{-2} = [(1+\lambda^2)/\pi^2]\langle 1s|er|2p\rangle^{-2} \quad (10)$$

so that the measured ratio $t_p/t_s = \lambda^2$ gives the admixed fraction $\lambda$ directly. The experimental values being $t_p = 253$ ns and $t_s = 570$ ns, $\lambda = 0.67$ is obtained (KCl).[4] This gives $\phi(q) = 0.39$ and $\tan(\phi(q)) = 0.42$ at the emitting bottom-well configuration $q = \{q_i\}$. From $E_{sp} = 0.12$ eV [5] we get $\sqrt{\sum_i(2G_iq_i+D_i)^2} = 0.287$ eV. We find $D = 0.166$ eV if $q \sim 0$.

Further on, we use $\varepsilon = 2.19$, $\alpha_0 = 2.3234$ Å, $V_M = 7.94$ eV, $r_0 = 3.147$ Å and the mixing constant equations $u_{x,x} = 2.235(V_M/r_0)/(\alpha r_0)$ and $v_{x,x} = -6.704(V_M/r_0)/(\alpha r_0)$ to compute $\alpha_{1s2p}r_0 = 2.032$, $G_{1s2p} = 2.775$ eV/Å and $\alpha_{2s2p}r_0 = 1.354$, $G_{2s2p} = -12.489$ eV/Å. The resulting Jahn-Teller energies $E_{JT} = G^2/2K$ are 2.81 eV and 56.8 eV, respectively, at $K \approx K_{T1u} = (3/2)(K_{LBM}/6) = 1.373$ eV/Å$^2$. Now since $4E_{JT}/E_{sp} \gg 1$ in both cases, it follows that the $T_{1u}$ mixing at the KCl F center is strong between either s state, on the one hand, and the p state, on the other.

Unfortunately, no equivalent lifetime data have been reported so far on the NaI F center but an estimate of $t_s$ is based on less accurate emission data. For this reason we will rather follow the other way round. From Table I we have $E_{sp}$ = 0.114 eV and assume G = 0.623 eV/Å, D = -0.114 eV, as estimated in Section 2. Next, for q ~ 0 we calculate f(q) = ¼ π = 0.785 and α = -0.414 from equations (7) & (8), respectively, so that $t_p / t_s = \alpha^2$ = 0.172. Now if $t_s$ ~ 20 ns [14] then $t_p$ ~ 3 ns is predicted by these data.

It may be profitable to speculate on the implications of the $T_{1u}$ coupling on the configurational structure of the F center. If the mixing is strong as it appears to be the case for 2s-2p, then the cavity will undergo pressure tending to push it along <111>. The resulting metastable cavity displacements along the body diagonals end up on the surface of a sphere of radius

$$q_0 = \sqrt{(2E_{JT}/K)}\sqrt{(1-\mu^2)}, \tag{11}$$

containing the undisturbed cavity, where

$$E_{JT} = G^2/2K \tag{12}$$

is the Jahn-Teller energy and $\mu = E_{sp}/4E_{JT}$. Due to these outward displacements the effective cavity radius will be expected to increase to $r_0+q_0$ and the F band peak energy to decrease following Mollwo-Ivey's law $E_F \sim 1/(r_0+q_0)^2$. This will bring about measurable red shifts of F bands in alkali halides with small cationic radii (low cation-to-anion radial ratios) which favor the $T_{1u}$ displacements.

We stress that the off-center sites are metastable, due to the off-on barrier which amounts to $E_{Boff-on} = (1-\mu)^2 E_{JT}$ in the symmetric D = 0 case. The trapped electron will be smeared in the extended cavity by virtue of a tunneling rotation of vibronically coupled F center between metastable sites at $r_0+q_0$ along nn body diagonals. These sites are separated from one another by intersite barriers of magnitude $E_{Brot} = I\omega_{ren}^2/8$ where $I = (3/2)Mq_0^2$ is the inertial moment, M is the nn cation mass, $\omega_{ren} = \omega_{bare}\sqrt{[(T_{113}-T_{111})/4G]q_0}$ is the renormalized mode frequency [8]. The smeared F center will have rotational energies in one of the tunneling bands which is but small with respect to the optical transition energies.

## 5. Nonradiative rates

Some time ago, De Matteis et al.[12] argued that a nonradiative deexcitation to the ground 1s-like state accounts for the some of the processes following the optical excitation of the NaBr F center [13-15]. In view of their conclusions we computed the de-excitation rate of an F center, following optical excitation to the 2s'- or 2p'-like vibronic mixtures, via a nonradiative transition to the 1s-like state. For that purpose we used the relevant weak-coupling equations of Part I with the entering parameters for NaI as given in Table I. The resulting deexcitation rates $k_{2p'-1s}(T)$ and $k_{2s'-1s}(T)$ were temperature dependent though too low. Otherwise, they were akin to the nonradiative interwell rate dependences $k_{2s'-2s'}(T)$ and $k_{2p'-2p'}(T)$, as in Figure 3.

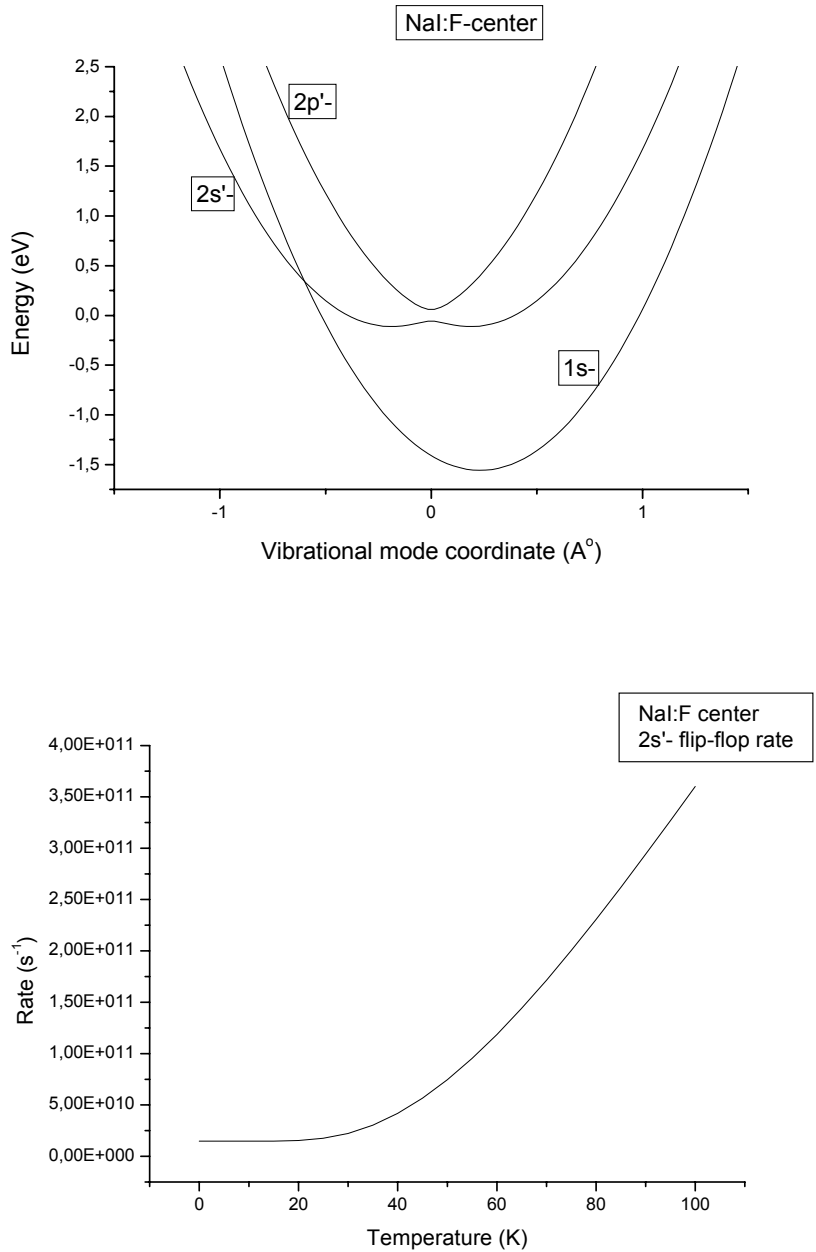

Figure 3:
Top: The configurational coordinate diagram of NaI F center states accounting for the pseudo-Jahn-Teller splittings in the excited electronic state. Two radiative deexcitation transitions are conceivable occurring from the bottoms on the lower 2s'- branch to the 1s- like ground state. Their energies are comparable with the ones listed in Table II. Bottom: The temperature dependence of the interwell flip-flop rate $k_{2s'-2s'}$ along the 2s'- adiabatic surface which is competitive to the nonradiative deexcitation rate $k_{2s'-1s}$ by prolonging the 2s'- lifetime. Note the protracted low-temperature or zero-point rate extending below 20 K.

For the sake of completeness, we calculated the interwell flip-flop rate $k_{2s'-2s'}(T)$ between the minima on 2s', the lower branch of the excited state adiabatic surface. The latter transition is essential as part of the relaxation following the one-phonon absorption of the F center. For that purpose, we now reproduce the relevant strong-coupling formulae for the transition probabilities.

At overbarrier levels, $E_n > E_C$, the electron-transfer term is

$$W_e(E_n) = 2[1-\exp(-2\pi\gamma_n)] / [2-\exp(-2\pi\gamma_n)]$$

while the configurational tunneling term is near 1. At subbarrier levels, $E_n < E_C$, the electron-transfer term is

$$W_e(E_n) = 2\pi\gamma_n^{2\gamma-1}\exp(-2\gamma_n)/[\Gamma(\gamma_n)]^2$$

while the configurational term is

$$W_L(E_n) = \pi\{[F_{nm}(\xi_0,\xi_C)]^2 / 2^{n+m}n!m!\}\exp(-Q^2/E_R\hbar\omega)\exp(-E_R/\hbar\omega)$$

where

$$F_{nm}(\xi_0,\xi_C) = \xi_0 H_n(\xi_C)H_m(\xi_C-\xi_0) - 2nH_{n-1}(\xi_C)H_m(\xi_C-\xi_0) + 2mH_n(\xi_C)H_{m-1}(\xi_C-\xi_0).$$

$\xi = \sqrt{(K/\hbar\omega)}q$ stands for the dimensionless mode coordinate, $\xi_0$ and $\xi_C$ are the final state equilibrium coordinate and the crossover coordinate, respectively, $H_{nm}(...)$ are Hermite's polynomials. The remaining symbols are defined in Part I. The interwell relaxation rate is composed by summing up the weighed transition probabilities:

$$k_{if}(T) = (\omega/\pi)\sinh(\hbar\omega/2k_BT)\sum_n W_e(E_n)W_L(E_n)\exp(-E_n/k_BT),$$

also explained therein, is shown in Figure 3.

### 5.1. Nonradiative deexcitation rate in F centered NaI

The 1s-2s mixing by $T_{1u}$ mode via pseudo-Renner effect, still hypothetical though, would have profound implications for the relaxation following the optical excitation of an F center. In as much as the 1s-2s mixing ratio in the 2s-like excited state may be expected to grow increasingly larger towards the 2s-1s crossover essential for the radiationless process, an extra nonradiative channel is suggested due to the increased adiabatic energy splitting. At the present stage, however, the equilibrium nonradiative deexcitation is rather not competitive to the intervibrational relaxation, due to the large number of phonons (~ 100) exchanged with the thermal bath during the relaxation. (See Table I for data on the electron binding energies.)

### 6. Conclusion

In Part I, we outlined the premises of a uni-mode approach to the non-radiative de-excitation of F centers in alkali halides. In view of experimental Raman data the single-frequency analysis has a better chance to working in crystals with low ratios of the cation to anion radii, such as NaI and NaBr [16]. In Part II we theorized over various mixing mechanisms of electronic states which might be found useful for applications to calculating the nonradiative deexcitation rate. In Part III we extended the method so as to incorporate more than one mode frequency. The vibrational frequency correspondence in the initial and final electronic states is essential for the horizontal transitions to occur between the initial and final state. Nevertheless the multimode coupling can be dealt with in cases where it allows for a single effective-frequency approach. Excited state 2s-2p mixing by $T_{1u}$-mode coupling is analyzed in detail with implications supporting Ham's conclusions. The non-radiative relaxation rates from 2p'- and 2s'- mixture states down to the 1s-like ground state are calculated and compared with the intervibrational relaxation rate.

Table I
Self-consistent semi-continuum potential calculations for NaI ($A_{1g}$ coupling)[a]

| State | Cavity Radius $r_0$ (Å) | Cavity Potential $V_0$ (eV) | Dielectric Constant $\varepsilon$ | Wavefunction Parameters $V=\kappa r_0$  $u=\alpha r_0$ | Normalized Constants $A$ (Å$^{-3/2}$)  $B$ |
|---|---|---|---|---|---|
| 1s- | 3.237 | 5.003 | 3.100 | 1.562   .985 | .201   .345 |
| 2s- | 3.237 | 4.924 | 2.965 | 1.870   .515 | .072   .127 |
| 2p- | 3.237 | 4.557 | 3.510 | 1.430   .435 | .042   .057 |

| State | Electron Energy $\langle\psi|H_e|\psi\rangle$ (eV) | Coupling Constant $\langle\psi|b(r)|\psi\rangle$ (eV/Å) | Vibrational Quantum $\hbar\omega_{vib}$ (meV) | Force Constant $K=M\omega^2$ (eV/Å$^2$) | Configurational Coordinates $q_{min}$   $q_{crossover}$  1s-  2s- (Å) |
|---|---|---|---|---|---|
| 1s- | -1.410 | 1.273 | 12.9 | 5.492 | -.232       1.028 |
| 2s- | -.257 | .151 | 12.9 | 5.492 | -.028  1.028 |
| 2p- | -.143 | -1.071 | 12.9 | 5.492 | .195  .541  .093 |

| State | Lattice Relaxation Energy $E_R$ (eV) 1s-   2s- | Crossover Energy Barrier $E_C$ (eV) 1s-   2s- | Electron Binding Energy $Q_\psi$ (eV) | Electron Energy Splitting $V_{eg}$ (eV) 1s- 2s- 2p- | Vibrational Splitting $\hbar\Delta\omega_{vib}$ (meV) |
|---|---|---|---|---|---|
| 1s- |           .114 |           3.06 | -1.557 |   .118   0 | $10^{-2}$ |
| 2s- | .114 | 3.06 | -.259 | .116      0 | $10^{-2}$ |
| 2p- | .500   .136 | .328   .028 | -.247 | 0    0 | $10^{-2}$ |

Table II
Optical energies of NaI F center
Experimental / Theoretical

|  |  |
|---|---|

| Optical Bands | | Apparent | | |
|---|---|---|---|---|
| Absorption | Emission | Ionization Barrier | | |
| $E_A$ (eV) | $E_{E1}$ (eV)  $E_{E2}$ | $E_B$ (eV) | | |
| 2.08 | .75       .56 | .03 | | |
| 1.98 | | 2s-1s | 2p-1s | 2p-2s |
| | | 1.20 | .76 | .03 |

Table III

$T_{1u}$ coupling parameters for NaI F center (see details in Refs. [8] and [10])

| Effective Force Constant K (eV/Å$^2$) | 1$^{st}$ order Mixing Constants $G_{1s2p}$ $G_{2s2p}$ (eV/Å) | 3$^{rd}$ order Mixing Constants $T^{111}_{1s-2p}$ $T^{111}_{2s-2p}$ $T^{112}_{1s-2p}$ $T^{112}_{2s-2p}$ (eV/Å$^3$) | Jahn-Teller Energy $E_{JT1s-2p}$ $E_{JT2s-2p}$ (eV) | Energy Gap $E_{1s-2p}$ $E_{2s-2p}$ (eV) | Mixing Strength[b] $4E_{JT}/E_{sp}$ | Interwell Barrier[c] $E_B$ (eV) |
|---|---|---|---|---|---|---|
| 5.492 | -1.636 | -0.074 | 0.244 | 1.153 | 0.845 | |
| 5.492 | -6.604 | -0.546 | 3.971 | 0.114 | 140 | 4 |
| 5.492 | | -1.684 | | | | |
| 5.492 | | -6.109 | | | | |

.
[a]Wave function Bohr radius $a_0$ = 3.10517 Å, Madelung's potential $V_M$ = 7.73 eV.
[b]Mixing processes of strength ~ 1 are referred to as "weak", while they are "strong" otherwise. [c]Symmetric barrier at D=0.

Values of the mixing constants, as calculated using Table I data are themselves listed in Table III. A problem arises related to the force constant in an effective single-mode interpretation of the data therein. Namely, while the frequency of the $T_{1u}$ mode is not available, we assume the TO-mode value from the Restrahlen peak: 0.0144 eV (NaI). (The LO frequency 0.0228 eV obtains through multiplying by $\sqrt{(\varepsilon_0/\varepsilon_\infty)}$). Further the equilibrium $T_{1u}$ coordinate should be close to the LBM (local breathing mode) coordinate for the 1s-like F state because of the balancing effect of the negative charge in the cavity in both cases. Consequently the partial reorganization energy of $T_{1u}$ will be small and the effective bi-mode frequency ω along the reaction coordinate should be close to $\omega_{LBM}$ = 0.0129eV, while the effective force constant K is close to $K_{LBM}$. We see that with K = $K_{LBM}$ the 1s-2p vibronic mixing is weak, while the 2s-2p mixing is strong, as in Figure 2.

Appendix I

Extended mixing Hamiltonian
AI.1. General secular equation

Considering the $T_{1u}$- mixing of F center states we account for the odd-mode symmetry in the light of the requirement that the complete Hamiltonian must be invariant to the respective point-group operations [16]. Concomitantly, mixing may be of an odd-number order when it comes to different-parity states and of an even-number order for states of the same parity. Accordingly we define

$$H = \tfrac{1}{2}(\sum_i \mathbf{P}_i^2/M_i + \sum_i K_i q_i^2) + E_{1s}|1s\rangle\langle 1s| + E_{2s}|2s\rangle\langle 2s| + \sum_i E_{2p}|2p_i\rangle\langle 2p_i|$$

$$+ \sum_i (G_{12i} q_i + D_{12i})(|1s\rangle\langle 2p_i| + |2p_i\rangle\langle 1s|)$$

$$+ \sum_i (G_{22i} q_i + \sum_{jk} T_{ijk} q_i q_j q_k + D_{22i})(|2s\rangle\langle 2p_i| + |2p_i\rangle\langle 2s|)$$

$$+ (C_{12} q^2 + D_{12})(|1s\rangle\langle 2s| + |2s\rangle\langle 1s|) + \sum_{ij}(C_{ij} q_i q_j + D_{ij})(|2p_i\rangle\langle 2p_j| + |2p_j\rangle\langle 2p_i|)$$

where $G_i$, $C_{ij}$ ($C_{12}$) and $T_{ijk}$ at $i,j,k = x,y,z$ are first-, second- and third- order coupling constants, respectively. By virtue of symmetry $b_i = b$, $c_{ij} = c_p$, $T_{ijk} = 0$ ($i \neq j \neq k \neq i$), $= T_b$

($i=j\neq k$), $= T_c$ ($i=j=k$). We next seek for the eigenvalues of the adiabatic Hamiltonian $H_{AD} \equiv H - \frac{1}{2}\sum_i P_i^2/M_i$ by means of the linear combination:

$$\psi(\mathbf{r},q_i) = c_1|1s\rangle + c_2|2s\rangle + c_x|2p_x\rangle + c_y|2p_y\rangle + c_z|2p_z\rangle$$

which leads to the following secular equation:

|  | 1s | 2s | $2p_x$ |
|---|---|---|---|
| 1s | $E-\frac{1}{2}\sum_i K_i q_i^2 - E_{1s}$ | $C_{12}q^2 + D_{12}$ | $G_{12x}q_x + D_{12x}$ |
| 2s | $C_{12}q^2 + D_{12}$ | $E-\frac{1}{2}\sum_i K_i q_i^2 - E_{2s}$ | $G_{22x}q_x + \sum_{jk} T_{xjk} q_x q_j q_k + D_{22x}$ |
| $2p_x$ | $G_{12x}q_x + D_{12x}$ | $G_{22x}q_x + \sum_{jk} T_{xjk} q_x q_j q_k + D_{22x}$ | $E-\frac{1}{2}\sum_i K_i q_i^2 - E_{2p}$ |
| $2p_y$ | $G_{12y}q_y + D_{12y}$ | $G_{22y}q_y + \sum_{jk} T_{yjk} q_y q_j q_k + D_{22y}$ | $C_{xy}q_xq_y + D_{xy}$ |
| $2p_z$ | $G_{12z}q_z + D_{12z}$ | $G_{22z}q_z + \sum_{jk} T_{zjk} q_z q_j q_k + D_{22z}$ | $C_{xz}q_xq_z + D_{xz}$ |

|  | $2p_y$ | $2p_z$ |  |
|---|---|---|---|
|  | $G_{12y}q_y + D_{12y}$ | $G_{12z}q_z + D_{12z}$ |  |
|  | $G_{22y}q_y + \sum_{jk} T_{yjk} q_y q_j q_k + D_{22y}$ | $G_{22z}q_z + \sum_{jk} T_{zjk} q_z q_j q_k + D_{22z}$ |  |
|  | $C_{yx}q_yq_x + D_{yx}$ | $C_{zx}q_zq_x + D_{zx}$ | $= 0$ |
|  | $E-\frac{1}{2}\sum_i K_i q_i^2 - E_{2p}$ | $C_{zy}q_zq_y + D_{zy}$ |  |
|  | $C_{yz}q_yq_z + D_{yz}$ | $E-\frac{1}{2}\sum_i K_i q_i^2 - E_{2p}$ |  |

Taking account of the symmetry of the coupling constants over x, y, and z, we have

|  | 1s | 2s | $2p_x$ |
|---|---|---|---|
| 1s | $E-\frac{1}{2}Kq^2 - E_{1s}$ | $C_{12}q^2 + D_{12}$ | $G_{12}q_x + D_{12}$ |
| 2s | $C_{12}q^2 + D_{12}$ | $E-\frac{1}{2}Kq^2 - E_{2s}$ | $G_{22}q_x + \sum_{jk} T_{xjk} q_x q_j q_k + D_{22}$ |
| $2p_x$ | $G_{12}q_x + D_{12}$ | $G_{22}q_x + \sum_{jk} T_{xjk} q_x q_j q_k + D_{22}$ | $E-\frac{1}{2}Kq^2 - E_{2p}$ |
| $2p_y$ | $G_{12}q_y + D_{12}$ | $G_{22}q_y + \sum_{jk} T_{yjk} q_y q_j q_k + D_{22}$ | $C_{pp}q_xq_y + D_{pp}$ |
| $2p_z$ | $G_{12}q_z + D_{12}$ | $G_{22}q_z + \sum_{jk} T_{zjk} q_z q_j q_k + D_{22}$ | $C_{pp}q_xq_z + D_{pp}$ |

|  | $2p_y$ | $2p_z$ |
|---|---|---|

$$\begin{array}{ll}
G_{12}q_y+D_{12} & G_{12}q_z+D_{12} \\
G_{22}q_y+\sum_{jk}T_{yjk}q_yq_jq_k+D_{22} & G_{22}q_z+\sum_{jk}T_{zjk}q_zq_jq_k+D_{22} \\
C_{pp}q_yq_x+D_{pp} & C_{pp}q_zq_x+D_{pp} \\
E-\tfrac{1}{2}Kq^2-E_{2p} & C_{pp}q_zq_y+D_{pp} \\
C_{pp}q_yq_z+D_{pp} & E-\tfrac{1}{2}Kq^2-E_{2p}
\end{array} = 0$$

## Appendix II

### The "Reflected diagram method" for calculating weak-coupling transition rates

Suppose we have a weak-coupling configurational diagram like any of the depicted pairs in Figure 1. Weak-coupling occurs when the crossover point is outside the frame formed between the two minima on the pair of diagrams, as between the ground state parabola 1s- and the excited state 2s- parabola in Figure 4. To calculate a weak-coupling rate, he first step is to reflect the excited state parabola from a perpendicular plane passing through the origin or any other symmetry center on the pair of diagrams. A reflected- image diagram $2s_r$-is thus formed. Denoting the relevant concentrations as they stand, we get $F_0 = [1s-] + [2s-] + [2s_r-]$ which is differentiated to give in steady state ($F_0$ is the total F-center concentration)

$$d[1s-]/dt = -q_F[1s-] + k_{2sr-1s}[2s_r-] = 0$$

$$d[2s-]/dt = +q_F[1s-] - k_{2s-2sr}[2s-] = 0$$

$$d[2s_r-]/dt = k_{2s-2sr}[2s-] - k_{2sr-1s}[2s_r-] = 0$$

where $q_F$ is the (two-) photon absorption rate. Inserting we get

$$F_0 = [1s-]\{1 + (q_F/k_{2s-2sr}) + (q_F/k_{2sr-1s})\}$$

The overall excited-state concentration therefore is

$$F^* = q_F F \tau_F = q_F [1s-] \{(1/k_{2s-2sr}) + (1/k_{2sr-1s})\}$$

where $\tau_F = (1/k_{2s-2sr}) + (1/k_{2sr-1s}) = 1/k_{2s-1s}$ is the F center nonradiative lifetime. The apparent 2s- to 1s- nonradiative deexcitation rate $k_{2s-1s}$ therefore reads:

$$k_{2s-1s} = k_{2s-2sr} k_{2sr-1s} / (k_{2s-2sr} + k_{2sr-1s}) \qquad (AII.1)$$

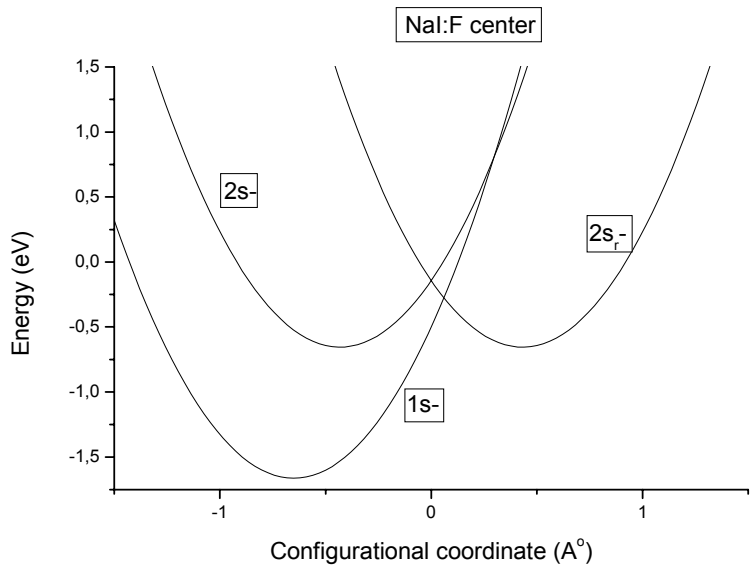

Figure 4:
A configurational coordinate diagram illustrating the reflected-diagram method for calculating the nonradiative deexcitation rate in a weak-coupling situation, e.g. the one formed between a ground-state parabola, say 1s-, and an excited-state parabola, say 2s-. The method is alternative to the related formulae in Part I. In that method, the excited-state curve is reflected from a perpendicular plane passing through the origin to obtain an image curve $2s_r$. The nonradiative rate is then calculated by equation (AII.1) using components derived for strong-coupling situations.